\documentclass[prl,preprintnumbers,superscriptaddress,showpacs,twocolumn,balancelastpage,floatfix]{revtex4}

\usepackage{amsmath,amssymb,amsfonts}
\usepackage{bm,graphicx,color}

\newcommand{\TR}{\text{Tr}}

\newcommand{\Vk}{{\bm k}}

\newcommand{\expval}[1]{\langle{#1}\rangle}

\newcommand{\ret}{{\text{R}}}
\newcommand{\adv}{{\text{A}}}

\newcommand{\les}{<}
\newcommand{\gtr}{>}

\newcommand{\teff}{T_\text{eff}}
\newcommand{\Etot}{\mathcal{E}}

\begin{document}

  \title{Damping of Bloch oscillations in the Hubbard model}

\author{Martin Eckstein}
\affiliation{Theoretical Physics, ETH Zurich, 8093 Zurich, Switzerland}
\author{Philipp Werner}
\affiliation{Theoretical Physics, ETH Zurich, 8093 Zurich, Switzerland}  
  
\date{\today}
\begin{abstract}
\noindent
Using nonequilibrium dynamical mean-field theory, we study the isolated Hubbard model in a static 
electric field in the limit of weak interactions. Linear response behavior is established at long times,
but only if the interaction exceeds a critical value, below which the system exhibits an AC-type 
response with Bloch oscillations. The transition from AC to DC response is defined in terms of the universal long-time 
behavior of the system, which does not depend on the initial condition. 
\end{abstract}

\pacs{71.10.Fd}

\maketitle

In the absence of scattering of charge carriers in a metal, a 
static
electric field results in undamped oscillations of the current, which are known as 
Bloch oscillations. The fate of these oscillations in the presence of strong inter-particle 
scattering is theoretically not well understood. Intuitively, one might expect them to get
damped until a direct current (DC) is established at long times, which would then 
be given by the linear or nonlinear DC response of the system. In the following we 
demonstrate that this intuitive picture is not true in general for a closed system%
: For the Hubbard model, we show that 
an electric field induces a DC response only if the inter-particle interaction exceeds 
a critical value.

Bloch oscillations are most easily understood in a simple tight-binding model.
For example, if a linear potential is added to a tight-binding chain with
lattice spacing $a$, the single-particle spectrum changes from a continuous 
energy band to an infinite set of levels at integer multiples of the potential 
difference $eaE$ between neighboring lattice sites (for a review, see 
Ref.~\cite{Glueck2002}). The eigenstates 
of this so-called Wannier-Stark ladder are localized on a length 
$l\propto 1/E$, and beating oscillations at the Bloch 
frequency $\omega_B=eaE/\hbar$ arise from  any  linear superposition
of those states. A direct experimental observation of Bloch oscillations in 
solids is hardly possible because extremely large fields are needed to make 
the period $2\pi/\omega_B$ short compared to typical scattering times. 
However, Bloch oscillations have been observed in semiconductor superlattices 
\cite{Leo1998}, and, within a well-controlled setup, using  ultracold atomic 
gases in optical lattices \cite{coldatom}.

Our initial question about the establishment of a DC regime becomes 
somehow trivial for a system that is coupled to a thermal bath. In this
case one will always get a finite current at long times, although for 
large fields  the magnitude of this current can exhibit an interesting 
dependence on the system-bath coupling \cite{Amaricci2011}.  A 
closed system, on the other hand, which is the appropriate representation 
for cold atoms in an optical lattice, cannot sustain a true steady state 
with nonzero current $j$ in a constant field, because the energy $\Etot$ 
always changes at a rate $\dot\Etot=Ej$ (e.g.,  Ref.~\cite{Mierzejewski}). 
So the question arises how one can possibly define  
a transition from an oscillating to a direct current in such a system. 
As it turns out, the answer to this is already the key for understanding 
the nature of the transition itself: 
While the true steady current is zero, the 
system establishes a universal relation between its thermodynamic 
quantities and the current well before the final state is reached, and 
it is by means of this universal behavior that one can clearly separate 
a linear response-like DC regime from an alternating current (AC) regime, 
in which the system exhibits Bloch oscillations at long times.

In this paper we investigate the AC/DC transition within the half-filled Hubbard model,
\begin{equation}
\label{hubbard}
H= \sum_{ij,\sigma=\uparrow,\downarrow} \!\!t_{ij}\, c_{i\sigma}^\dagger c_{j\sigma}
+ U
\sum_{i}
\big(n_{i\uparrow}\!-\!\tfrac12\big)
\big(n_{i\downarrow}\!-\!\tfrac12\big),  
\end{equation}
which describes fermionic particles that can hop between the sites of a crystal 
lattice (with hopping amplitude $t_{ij}$) and interact with each other through a local 
Coulomb repulsion $U$. We will characterize the zero-current final state 
(which still feels the presence of both electric field and interaction), 
and demonstrate the existence of an AC/DC transition at an interaction
$U>0$. The results fit well into the picture established by a number of recent investigations on the topic. Exact 
diagonalization of the Bose Hubbard model shows a qualitative change of the 
many-body spectrum with increasing electric field \cite{Buchleitner2003}, and  
for spinless fermions, Bloch oscillations are observed in an integrable version 
of the model, while a nonintegrable version shows overdamped behavior 
\cite{Mierzejewski}. In the infinite-dimensional Falicov-Kimball model oscillations 
are damped \cite{Freericks2006a}, but the relaxation to the steady behavior
is still not fully resolved there. 
Moreover, our findings 
link the damping of Bloch oscillations to 
the more general question how a closed 
system relaxes to a well-defined state. This question has been intensively 
discussed recently, 
in order to understand
the thermalization of isolated many-body systems \cite{quenchreview}.

  
  We solve the dynamics of the Hubbard model using the dynamical mean-field theory 
  (DMFT) \cite{Georges1996} in its nonequilibrium variant \cite{Schmidt2002}. The electric 
  field is treated in a gauge with zero scalar potential and time-dependent vector 
  potential, $\bm{E}=-\frac{1}{c} \partial_t \bm{A}$. The latter  enters the Hamiltonian 
  (\ref{hubbard}) via a Peierls substitution, i.e, a time-dependent shift of the band energy 
  $\epsilon(\Vk)  \to \epsilon(\Vk- \bm{A})$. We choose the field along the 
  $(1,1,\ldots ,1)$-direction in the infinite-dimensional hypercubic lattice 
  with a Gaussian density of states $\rho(\epsilon)$ $=$ $e^{-\epsilon^2}/\sqrt{\pi}$. The unit 
  of energy is given by the variance  $W$ of the density of states (the bandwidth), 
  time is measured in units of $\hbar/W$, and the unit of the electric field is 
  given by $W/ea$, where $-e$ is the electronic charge and $a$ is the lattice spacing. 
  The DMFT equations for this setup have been discussed in detail in Ref.~\cite{Freericks2006a},
  and our precise implementation is given in Ref.~\cite{Eckstein2011}. Because we are interested in the regime 
  of weak coupling, we use iterated perturbation theory (IPT) \cite{Georges1996} to solve the effective
  impurity problem of DMFT. For nonequilibrium, IPT can work very well for small $U$ in 
  spite of the fact that it is not conserving, but it breaks down rather abruptly if $U$ is too 
  large \cite{Eckstein2009a}. To validate our results, we have performed Monte Carlo 
  simulations \cite{Werner2009}, which reproduce the AC/DC transition, but do not allow a systematic analysis 
  of the long-time behavior.

  If not stated otherwise, the results below show the time-evolution of the Hubbard model 
  in a constant electric field, starting from the free Fermi sea at $t<0$. Later we investigate various 
  other initial states and switch-on procedures of the field in order to show that the conclusions
  of the paper do dot depend on them. To characterize the time-evolving state we compute the 
  current
  $j(t)=\sum_\Vk \expval{c_\Vk^\dagger(t) c_\Vk(t)} \partial_\Vk \epsilon_\Vk$
  and the local spectral function (which is  gauge-independent \cite{Davies1988})
  \begin{equation}
  \label{aret}
  A(\omega,t) = - \frac{1}{\pi}\text {Im} \int_0^\infty ds\, G^\ret(t+s,t) e^{i\omega s},
  \end{equation}
  where $G^\ret(t,t')=-i\Theta(t-t')\expval{\{c(t),c^\dagger(t')\}_+}$ is the retarded Green function.
  For $U=0$,  $A(\omega)$  resembles the Wannier-Stark ladder,
   $A(\omega) = \sum_m \delta(\omega - m\omega_B) w_m$,
  where the weights 
  $w_m$ are given by the amplitudes of the Wannier-Stark  states which are 
  localized at sites with a potential energy difference  $m\hbar\omega_B$ \cite{Davies1988}.


\begin{figure}[h]
\centerline{\includegraphics[clip=true,width=0.95\columnwidth]{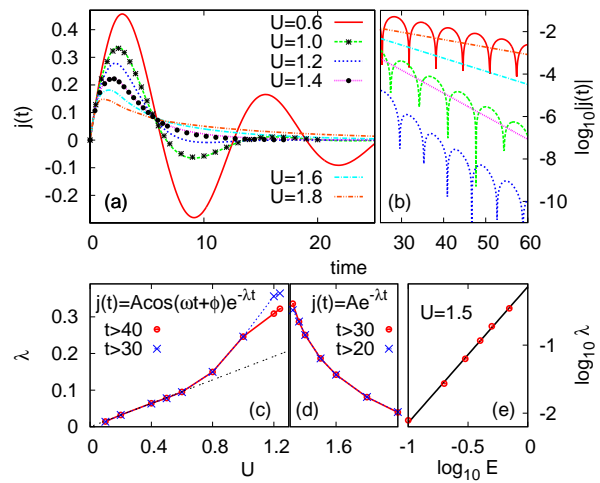}}
\caption{
(a) Current $j(t)$ for $E=0.5$ and various values of $U$. For $t<0$, the state is 
the noninteracting Fermi sea at $\beta=10$. Lines: IPT, Symbols (for $U=1.0$ and $U=1.4$): 
QMC. (b) Same parameters as (a), on a logarithmic scale. (c,d) Damping rate
$\lambda(U)$ for $E=0.5$, obtained from fitting $j(t)$ with a damped oscillator (c) and an exponential
decay (d), respectively. Each fit is computed for two time intervals to estimate the influence 
of the initial transients. (e) Damping rate $\lambda(E)$ for the DC regime ($U=1.5$,  exponential 
fit of $j(t)$). The line corresponds to $\lambda =\sigma_\infty/c_\infty E^2$,
where $\sigma_\infty=0.4172$ and $c_\infty=0.122$ have been computed for $U=1.5$ (see text). }
\label{fig-curr-e075-all}
\end{figure}

{\bf Results} ---
Figure \ref{fig-curr-e075-all}a and b show the time-dependent current after an electric
field $E=0.5$ is suddenly turned on in the Hubbard model. With increasing interaction, 
the evolution of the current changes from damped Bloch oscillations (AC regime) to 
a monotonously decreasing current (DC regime), which is best visible
on a logarithmic scale (Fig.~\ref{fig-curr-e075-all}b). For a quantitative characterization 
of the behavior we fit the data  in the AC and DC regimes  at long times with a
damped oscillation $j(t) = A \cos(\omega t + \phi) \exp(-\lambda t)$ and an exponential 
decay $j(t) = A \exp(-\lambda t)$, respectively (Fig.~\ref{fig-curr-e075-all}c and d). The 
fits work well everywhere except close to the transition,  where it apparently 
takes longer time until initial transients decay and a simple relaxation behavior is established 
(this will be discussed below).

\begin{figure}[h]
\centerline{\includegraphics[clip=true,width=1.0\columnwidth]{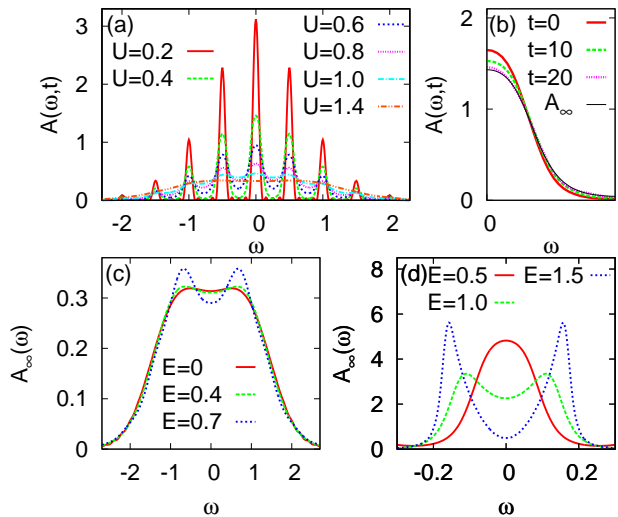}}
\caption{(a) Retarded spectrum, Eq.~(\ref{aret}), for the same parameters as in 
Fig.~\ref{fig-curr-e075-all} ($E=0.5$, $\beta=10$). (b) $A(\omega,t)$ in the central
frequency range for $U=0.4$ and various times, compared to the final state spectrum
$A_\infty(\omega;U,E)$. (c) The final state spectrum $A_\infty(\omega;U,E)$ in the 
DC regime ($U=1.5$). (d) The central Wannier-Stark peak of $A_\infty(\omega;U,E)$ 
in the oscillating regime ($U=0.4$), normalized to unit area.}
\label{fig-Aret}
\end{figure}

In the AC regime, the decay rate $\lambda(U)$ increases linearly up to $U \approx E$, 
where it exhibits a kink and starts to rise more rapidly (Fig.~\ref{fig-curr-e075-all}c). 
This result can be understood within the Wannier-Stark picture: For the given geometry, 
the energy levels of the tight-binding 
model with linear potential are given by integer multiples $m \hbar \omega_B$ of the Bloch 
frequency, and each level is highly degenerate due to the translational invariance of the 
system transverse to the field. Any interaction  $U \ll E$ will lift this degeneracy and 
lead to bands of width proportional to the matrix elements of the interaction operator in 
the manifold of Wannier-Stark states belonging to {\em one} energy. This splitting then 
leads to a dephasing of the oscillations at a rate proportional to $U$, and the kink can 
be associated with the fact that only for $U \gtrsim E$ scattering between Wannier Stark 
states with different $m$ becomes effective. 
The argument is supported by the behavior of the spectral function
(Fig.~\ref{fig-Aret}a). For $U\lesssim E$, we find that 
$A(\omega,t\to\infty)$ consists of well separated peaks with spacing $E$, whose 
weight is approximately given by the weight of the delta-peaks in the noninteracting 
spectrum of the Wannier Stark ladder. The gaps start to be filled for $U\gtrsim E$. 
Note that this crossover is not related to the transition between AC and DC 
regimes, which occurs only at larger values of $U$.

\begin{figure}[h]
\centerline{
\includegraphics[clip=true,width=1\columnwidth]{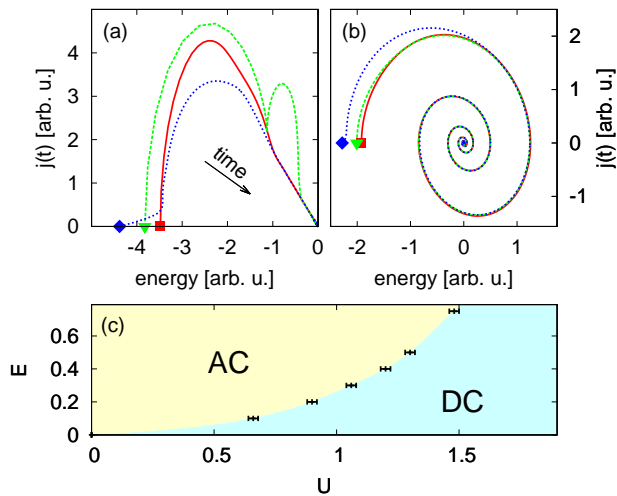}}
\caption{
(a) $j(t)$ plotted against the total energy $\Etot(t)$ at $E=0.6$, $U=1.8$
for various initial states: Sudden turn on of the field starting, from the noninteracting 
Fermi sea at $\beta=9$ (square symbol), gradual turn on of the field, $E(t)=E t/t_0$,
over a time-interval $t_0=10$ (diamond), and a sudden turn-on of the field starting 
from interacting thermal equilibrium at $\beta=5$ and $U=1.8$ (triangle). For each 
curve, $j$ and $\Etot$ have been multiplied by a single scaling factor. (b) Same 
as (a), for $U=0.6$ (c) $E-U$ phasediagram, showing regions where the long-time 
behavior is governed by oscillations (AC) or a direct current (DC). }
\label{fig-pd}
\end{figure}

In the DC regime, the decay rate $\lambda(U)$ decreases with the interaction 
(Fig.~\ref{fig-curr-e075-all}d). A simple explanation of the exponential decay of the current in this regime 
is possible in the limit of 
small $E$: Because the system is not coupled to a reservoir, the total energy $\Etot$ 
increases a the rate $\dot \Etot(t)  = E j(t)$ \cite{Mierzejewski}. 
The most simple assumption that one can make to account for this effect
is that the system rapidly thermalizes, 
such that its state is a thermal equilibrium 
state with temperature $\teff(t)$ and energy  $\Etot(t) = \TR[e^{-H/\teff(t)}H]/Z$ at any given 
time $t$. (From a Boltzmann equation, the thermalization time would be expected to be 
$\propto U^{-2}$.) The current at small $E$ is then given by the linear response value  
$j =\sigma_\text{dc} E$. At long times, the system approaches $\teff\to\infty$, and 
both $\sigma_\text{dc}$ and $\Etot$ 
are asymptotically given by the leading terms of their high-temperature expansion,
$\sigma_\text{dc} 
\sim \frac{\sigma_\infty }{\teff}$,  $\Etot \sim -\frac{c_\infty}{\teff}$. Hence, energy and current 
obey a linear relation
\begin{align}
\label{jetot}
j(t) \sim - E (\sigma_\infty / c_\infty)\, \Etot(t).
\end{align}
If this is inserted back into the exact relation $\dot \Etot  = E j(t)$, one finds that the current 
exhibits an exponential decay with rate
$\lambda \sim \sigma_\infty/c_\infty E^2$  for $E\to 0$. 
As a numerical check in the present case we verify the linear relation between
current and total energy at long times [Eq.~(\ref{jetot})] by plotting $j(t)$ against 
$\Etot(t)$ in Fig.~\ref{fig-pd}a. Also the $E^2$ dependence of $\lambda$ is
confirmed by our numerical results (Fig.~\ref{fig-curr-e075-all}e), where the 
coefficients $c_\infty$ and $\sigma_\infty$ are obtained by a solution of the 
DMFT equations in thermal equilibrium for $\beta\to 0$ (using IPT).
Interestingly, an analogous argument holds for a nonintegrable model of 
spinless Fermions \cite{Mierzejewski}. In contrast, rapid thermalization 
is impossible in the Mott insulating phase of the Hubbard model, such that 
a steady current can exist on rather long times \cite{Eckstein2010c}.

{\bf Steady state} ---
Both for the AC  and the DC regimes we find that the system ultimately approaches a 
peculiar steady state which carries no current in spite of the electric field. To obtain an 
understanding of this state, we start from the limit of infinite temperature, which is 
the only equilibrium state with zero conductivity. In equilibrium, the Green functions  
$G^\gtr(t,t')=-i\expval{c(t) c^\dagger(t') }$ and $G^\les(t,t')=i\expval{ c^\dagger(t')c(t)}$
are related by the fundamental relation $G^\les (\omega) =  - e^{\beta\omega} 
G^\gtr (\omega)$, such that one has
\begin{equation}
\label{tinf}
G^\les(t,t') = -G^\gtr(t,t') = \tfrac12 [ G^\ret(t,t')  - G^\adv(t,t')]
\end{equation}
at $\beta=0$. This ansatz, which treats quantum mechanical creation and annihilation 
operators as commuting objects, can readily be used as the definition of a generalized
infinite temperature state at nonzero $E$: It turns out that there is a unique steady-state 
solution $G^\ret(t,t')\equiv g_\infty(t-t')$ of the DMFT equations which satisfies 
Eq.~(\ref{tinf}): If Eq.~(\ref{tinf}) is enforced, IPT diagrams for the retarded self-energy 
can be expressed in terms of retarded Green functions only, in contrast to a general 
state, where they depend on the occupation functions, $G^\les(t,t')$ and  $G^\gtr(t,t')$. 
Hence, DMFT provides a closed set of equations for the the spectral (retarded) 
components of the Keldysh Green functions, which can be solved starting from the 
initial condition $G^\ret(t,t)=-i$. 
In Fig.~\ref{fig-Aret}b  we show that the spectral function (\ref{aret}) 
approaches $A_\infty(\omega) = -1/\pi \text{Im}g_\infty(\omega)$ 
for long times, which provides evidence that the state 
of the Hubbard model in a field at $t\to\infty$ 
is indeed  characterized by the ansatz (\ref{tinf}).  
In spite of its strong excitation, this state is still strongly influenced by the field,
both in the AC and DC regimes (Figs.~\ref{fig-Aret}c and d).
In particular, the Hubbard bands are enhanced in the presence of the field. 
An explanation for this fact could be that for the given geometry hopping 
between sites on one equipotential surface is possible only by a second 
order process via a site at potential difference $Eae$, so the bandwidth is 
effectively reduced to $W^2/Eae$ in the limit of strong fields.


{\bf The transition} ---
The current traces in Fig.~\ref{fig-curr-e075-all}b show that the switch from oscillating 
to plain exponential decay in the long-time behavior defines a sharp transition line 
$U_\text{ACDC}(E)$ in the $E$-$U$ diagram (Fig.~\ref{fig-pd}c). But how does this line depend
on specifics of the system, such as initial conditions, or the way in which the field is turned on? 
Figure \ref{fig-pd}a and b show plots of the current against the total energy for one set of 
parameters in the AC and  DC regime, respectively. By multiplication of $\Etot$ and $j$ 
with a single scaling factor, all curves for fixed $U$ and $E$ collapse to a unique path 
in the long-time limit.  This reveals the remarkable fact that the system follows a universal long-time 
behavior well before it reaches the final zero-current state discussed above. 
For small fields, this universal long-time behavior is precisely given by linear response 
theory [cf.~Eq.~(\ref{jetot})], but we can now see that the isolated Hubbard model 
actually follows this behavior only if the interaction exceeds the critical value 
$U_\text{ACDC}$.   

A universal long-time behavior naturally arises if the time-evolution for $t\to\infty $
can be described 
in terms of a linear equation for some reduced dynamical quantities $y(t)$. 
An example would be a Boltzmann equation, in which $y(t)$ are densities of relevant 
modes. If the equation is linearized close to a steady-state 
solution, the resulting linear equation has a number of exponentially decaying
eigenmodes, of which  the slowest survives at long times 
(with a single weight factor determined by the initial condition). A dynamical transition 
then occurs when
the relaxation times for two such qualitatively 
different solutions cross as a control-parameter is changed. The fact that the decay 
rate $\lambda(U)$ increases towards the transition both in the AC and the DC 
regime is consistent with this interpretation (\ref{fig-curr-e075-all}c and d).
Furthermore, close to the transition our data cannot be fit well with a simple relaxation
law, since they look more like a superposition of oscillating and decaying terms which 
are hard to separate. However, the derivation of a linearized
dynamical equation remains a unresolved issue for the present model. A starting point 
would be to linearize the exact time-evolution given by the Dyson equation around the 
final state given by the ansatz (\ref{tinf}), but such a calculation seems rather involved 
due to the time-dependence of the gauge dependent $\Vk$-resolved Green function 
in this state.

{\bf Conclusion} ---
In this paper we have studied the Hubbard model at weak $U$ in a static electric field $E$. 
In spite of the fact that the system is not coupled to a thermal reservoir, a DC response is 
established at long times. However, this holds only if the interaction exceeds a critical value, 
below which the system exhibits an AC-type response with Bloch oscillations. This AC/DC 
transition is defined by the long-time behavior of the system, which does no longer 
depend on the initial condition. Furthermore, we have related the damping rate of the 
Bloch oscillations to the destruction of the Wannier-Stark ladder, and we
have provided an understanding of the zero-current final state of the closed system 
in terms of a generalized infinite temperature state. Our results may be tested in 
experiments with ultracold atoms in optical lattices. Beyond this, we believe that a 
detailed understanding of the response of an isolated system is important in order to 
contrast studies of nonlinear transport in solid state bulk systems, where the precise 
form of the damping mechanism is often not known.

We gratefully thank M.~Kollar, A.~Lichtenstein,  M.~Mierzejewski, T.~Oka, P.~Prelov\v{s}ek, L.~Tarruell, 
N.~Tsuji, and A.~Zhura for useful discussions, and  A.~Lichtenstein for motivating this work.
Numerical calculations were run on the Brutus cluster at ETH Zurich. 
We acknowledge support from the Swiss National Science Foundation (Grant PP002-118866).

 \end{document}